\documentclass[aps,pra,twocolumn,10pt,bibnotes,raggedbottom,showpacs,floatfix]{revtex4-1}
\usepackage{amsmath,amsfonts,amssymb,amsthm,graphics}
\usepackage[next]{inputenc}
\usepackage[dvips]{epsfig}
\usepackage[colorlinks=true,citecolor=blue,linkcolor=blue]{hyperref}
\usepackage{bbm}
\usepackage{bbold}
\usepackage{booktabs}
\usepackage{multirow}
\usepackage{hhline}
\usepackage{amssymb}
\usepackage{comment}
\usepackage{float}
\usepackage{bm}
\usepackage[dvipsnames]{xcolor}
\usepackage{graphicx,latexsym}
\usepackage{epstopdf}

\begin{document}
\renewcommand{\vec}{\mathbf}
\renewcommand{\Re}{\mathop{\mathrm{Re}}\nolimits}
\renewcommand{\Im}{\mathop{\mathrm{Im}}\nolimits}

\title{Measurement induced dynamics and stabilization of spinor condensate domain walls}
\author{Hilary M. Hurst}
\author{I. B. Spielman}
\affiliation{Joint Quantum Institute, National Institute of Standards and Technology, and University of Maryland, Gaithersburg, Maryland, 20899, USA}

\begin{abstract}
Weakly measuring many-body systems and allowing for feedback in real-time can simultaneously create and measure new phenomena in quantum systems. We theoretically study the dynamics of a continuously measured two-component Bose-Einstein condensate (BEC) potentially containing a domain wall, and focus on the trade-off between usable information obtained from measurement and quantum backaction. Each weakly measured system yields a measurement record from which we extract real-time dynamics of the domain wall. We show that quantum backaction due to measurement causes two primary effects: domain wall diffusion and overall heating. The system dynamics and signal-to-noise ratio depend on the choice of measurement observable. We propose a feedback protocol to dynamically create a stable domain wall in the regime where domain walls are unstable, giving a prototype example of Hamiltonian engineering using measurement and feedback. 
\end{abstract}

\maketitle
\section{Introduction}
Understanding system-reservoir dynamics in many body physics is a new frontier. An external bath can be thought of as a `measurement reservoir' from which the environment extracts information about the system~\cite{Gardiner2004, Daley2014}. From this perspective, minimally destructive (i.e. backaction-limited) measurements constitute a controlled reservoir that also provides a time-resolved but noisy record of system evolution~\cite{Andrews1996, Andrews1997, Higbie2005, Liu2009}. Weak measurement has long been implemented in quantum-optical systems to monitor and control nearly-pure quantum states~\cite{Wiseman2009, Daley2014}, or in spin ensembles to create squeezed states~\cite{Wineland1992, Kuzmich2000, Schleier2010}. Extending this understanding to interacting many-body systems opens the door to measurement and quantum control of new, otherwise inaccessible strongly-correlated matter.

We theoretically investigate weakly measured spinor Bose-Einstein condensates (BECs), an experimentally accessible system for which closed system dynamics are well known~\cite{Stamper2013}. We explore measurement protocols sensitive to domain walls in two-component BECs, where the resulting measurement record tracks the domain wall over time. Furthermore, we show that classical feedback based upon the measurement record can create and stabilize domain walls. This process of `stochastic stabilization' via feedback from a noisy environment occurs in many other contexts, such as cell differentiation in biology whereby environmental noise can stabilize specific cell characteristics~\cite{Losick2008, Weber2013}.

Spinor condensates are predicted to host exotic spin texture defects such as skyrmions and non-abelian vortices~\cite{Kobayashi2009, Coen2001, Stamper2013, Matthews1999, Anderson2001, Ohberg2001, Ieda2004}. These defects interact with local exictations and undergo diffusion;  in real systems the excitations further destabilize many exotic spin textures~\cite{McDonald2016, Efimkin2016, Aycock2017, Hurst2017}. Stabilizing non-abelian excitations using current techniques has proven difficult, but might be possible using weak measurement and feedback, similar to our proposed approach for stabilizing a domain wall.

Domain walls in two-component BECs provide a test platform to understand the effects of repeated weak measurement on the stability and dynamics of topological defects. By combining quantum trajectory techniques (for open-system physics~\cite{Carmichael1993, Smith2013}) with Gross-Pitaevskii simulations (for closed system dynamics~\cite{Blakie2008,Symes2016}), we study the interplay of measurement, coherent evolution, and classical feedback. We propose two measurement protocols sensitive to the domain wall position and find that the choice of measurement observable affects both the heating rate and the dissipative dynamics of the domain wall.

\begin{figure}[t!]
\includegraphics{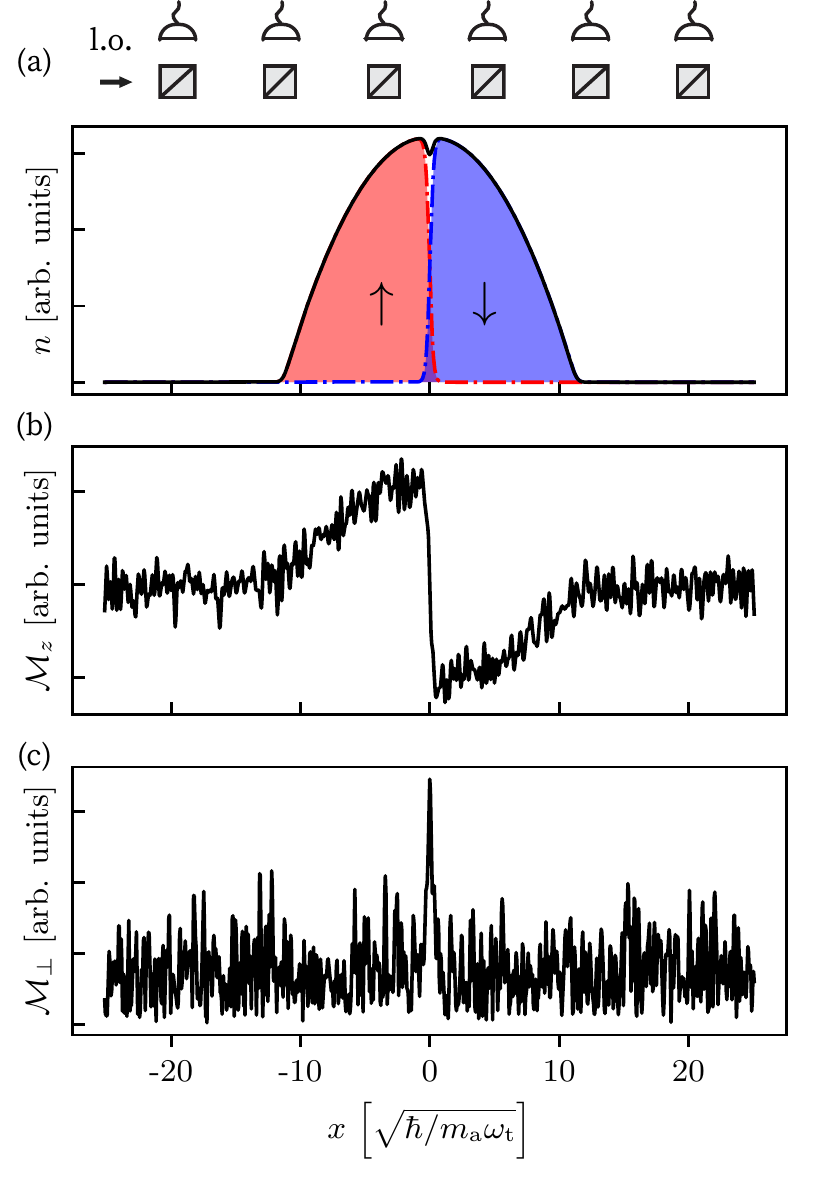}
\caption{(a)~Computed ground state system with a single domain wall and schematic illustration of phase contrast imaging layout. The system is weakly coupled to an array of homodyne detectors, where l.o.~indicates a strong local oscillator.  The BEC is phase separated into spin up (red/left) and spin down (blue/right); the black line indicates total density. (b-c)~Measurement outcome $\mathcal{M}$ of a single weak measurement with strength $\varphi=0.1$ of~(b)~$\mathcal{M}_z$ and ~(c)~$\mathcal{M}_\perp$ (defined in text).\label{Fig:1}}
\end{figure}

\section{Model}
\subsection{Measurement} We model spin-resolved dispersive imaging of a quasi-one-dimensional (1D) multicomponent condensate along $\vec{e}_x$ which interacts with a brief pulse of far detuned laser light of wavelength $\lambda$ and duration $\delta t$ traveling along $\vec{e}_z$~\cite{Shin2006, Gajdacz2013}. Here, the condensate is the system, and the light pulse is the `environment',  which is then subject to strong quantum measurement. We describe the optical field by the spatial mode basis $\sum_n\chi^*_{n}(z)\hat{a}_{nj}^\dagger$ where $\hat{a}^{\dagger}_{nj}$ describes the creation of a photon at $x_j$ (along the long axis of the 1D BEC)  in spatial mode $n$ and $\chi_{n}^*(z)$ is a normalized mode function (along the direction of the probe's propagation). We model the incoming probe beam as a coherent state with amplitude $|\alpha|$ and phase $\phi = \pi/2$ in a single spatial mode $\chi_0(z) = (c\delta t)^{-1/2}$, where $c$ is the speed of light. 

Atoms interact locally with the light via an interaction Hamiltonian described by a spin-dependent ac Stark shift~\cite{Brion2007},
\begin{equation}
\hat{H}^{\rm SR}_{\vec{r}j} =  \frac{\hbar\gamma}{c\delta t}\hat{S}_{\textbf{r}j} \otimes \hat{n}_j, 
\label{Eqn:HSR}
\end{equation}
where the reservoir operator $\hat{n}_j =\hat{a}^\dagger_j\hat{a}_j$ counts the photon number at $x_j$. The system operators $\hat{S}_{\textbf{r}j} =  \hat{b}^\dagger_{\sigma j}\left[\boldsymbol{\tau}\cdot\vec{r}\right]_{\sigma\sigma'}\hat{b}_{\sigma' j}$ measure the spin in the direction $\textbf{r}$, where $\hat{b}^\dagger_{\sigma j}$ describes the creation of an atom of spin $\sigma \in \left\{ \uparrow,\downarrow \right\}$ at $x_j$ and $\boldsymbol{\tau} = (\tau_x, \tau_y, \tau_z)$ is the vector of Pauli matrices. The system-reservoir interaction strength $\gamma$ is set by the atomic transition strength and the detuning from resonance.

Just prior to measurement, the system and reservoir mode evolve together for the pulse time $\delta t$ under the interaction unitary $\hat{U}_{\vec{r}} =  \exp\left[-i\varphi\sum_j\hat{S}_{\vec{r}j}\otimes \hat{Q}_j\right]$, which is a local displacement operator for the $\hat{X}_j$ quadrature of the optical field at $x_j$ where $\varphi =  \sqrt{2}\gamma|\alpha|/c$ is a small dimensionless parameter and $\left[\hat{X}_j,\hat{Q}_{j'}\right] = i\delta_{jj'}$~\footnote{See Supplemental Material}. The outcome of a single measurement for the full detector array is
\begin{equation}
\mathcal{M}_{\vec{r}}(x_j) = \langle \hat{S}_{\vec{r}j}\rangle + \frac{\vec{m}(x_j)}{\varphi}, \label{Eqn:Mresult}
\end{equation}
where $\vec{m}(x_j) $ is a vector describing quantum projection noise with momentum-space Gaussian statistics $\overline{\tilde{\vec{m}}_k} = 0$ and $\overline{\tilde{\vec{m}}_k\tilde{\vec{m}}_{k'}}  = \delta_{kk'}\Theta\left(|k|-k_c\right)/2$, where $\tilde{\vec{m}}_k$ denotes the Fourier transform of $\vec{m}(x_j)$, $\Theta$ is the Heaviside function, and $k_{\rm c} = 2\pi/\lambda$ denotes a momentum cutoff due to finite resolution. The momentum cutoff is implemented to account for the fact that the environment can only resolve information within a finite length scale $\lambda$.

A measurement with outcome $\mathcal{M}_{\vec{r}}(x_j)$ transforms the system wavefunction to $|\Psi_{|\vec{m}}\rangle = \hat{\mathcal{K}}_{\vec{r}|\vec{m}}|\Psi\rangle$ where $|\Psi\rangle$ is the system state before measurement and 
\begin{equation}
\hat{\mathcal{K}}_{\vec{r}|\vec{m}} \approx 1+\sum_j\varphi m_j\delta\hat{S}_{\vec{r}j}-\frac{\varphi^2}{4}\frac{k_c}{k_M}\left(\delta\hat{S}_{\vec{r}j}\right)^2\label{Eqn:KrausOp}
\end{equation}
is a Kraus operator corresponding to a global measurement of $\hat{S}_{\vec{r}}$, where $k_M = \pi/\Delta x$ is the maximum momentum in the simulation for grid spacing $\Delta x$ and $\delta\hat{S}_{\vec{r}j}  = \hat{S}_{\vec{r}j} - \langle\hat{S}_{\vec{r}j}\rangle$.

\subsection{System Dynamics} We describe the condensate in the mean-field approximation by a complex order parameter $\Psi_j = (\psi_{\uparrow j}, \psi_{\downarrow j})^T$ where $\psi_{\sigma j}$ is the coherent state amplitude of each spin (or pseudospin) component $\sigma \in \left\lbrace \uparrow, \downarrow\right\rbrace$ at $x_j$. The closed system evolves under the Gross-Pitaevskii equation (GPE)
\begin{equation}
i\hbar\partial_t\Psi_j= \left[\hat{H}_0+ u_0n_j\right]\Psi_j + u_2S_{zj}\tau^z\Psi_j, 
\label{Eqn:GPE}
\end{equation}
where $\hat{H}_0 = \hat{p}^2/2m_{\rm a} + m_{\rm a}\omega_{\rm t}^2x_j^2/2$ is the single particle Hamiltonian for atoms of mass $m_{\rm a}$ in a harmonic trap with frequency $\omega_{\rm t}$, $n_j = |\psi_{\uparrow j}|^2 +  |\psi_{\downarrow_j}|^2$ is the atom number at site $j$, and $S_{zj} = |\psi_{\uparrow j}|^2 -  |\psi_{\downarrow j}|^2$ is the atom number difference (magnetization) at site $j$. We work in units defined by the trap with $t \rightarrow t/\omega_{\rm t}$ and $x_j\rightarrow x_j\sqrt{\hbar/m_{\rm a}\omega_{\rm t}}$, and the wavefunction is normalized to the total number of atoms, $N = \sum_j n_j$. The spin-independent and spin-dependent interaction strengths $u_0 = 2\pi\hbar^2(a+a_{\uparrow\downarrow})/m_{\rm a}\Delta x$ and $u_2 = 2\pi\hbar^2(a-a_{\uparrow\downarrow})/m_{\rm a}\Delta x$ derive from the 1D intraspin and interspin scattering lengths $a$ and $a_{\uparrow\downarrow}$~\cite{Ho1998, Ohmi1998}. We fix the total atom number to be $N = 10^4$, and use $u_0\Delta x = 0.1$ and $u_2 = \pm 0.05u_0$, numbers which are representative of alkali atoms. For $u_2 < 0$ domain walls are stable, while for $u_2 > 0$ domain walls are unstable. The initial condition for all measurement simulations is the ground state of the GPE found by imaginary time evolution~\cite{Note1}. 

We calculate the Kraus operator's impact on the initial coherent state by assuming the system is well-described by a new mean-field coherent state after measurement, conditioned on the measurement result~\cite{Szigeti2009, Hush2013, Ilo2014, Wade2015}. To order $\varphi^2$ the coherent state
\begin{equation}
\Psi_{j|\vec{m}} = \left(1-\frac{\varphi^2}{4}\frac{k_c}{k_M}\right)\mathbb{1}\Psi_j + \varphi m_j\left[\boldsymbol{\tau}\cdot\vec{r}\right]\Psi_j
\label{Eqn:update}
\end{equation}
maximally overlaps with $\hat{\mathcal{K}}_{\vec{r}|\vec{m}}|\Psi\rangle$, thereby defining the updated coherent state. We numerically implement Eq.~\eqref{Eqn:GPE} using a second-order symplectic integration method~\cite{Symes2016}. For each measurement, we apply Eq.~\eqref{Eqn:update} to the wavefunction with a randomly generated noise vector $\vec{m}(x_j)$ leading to a stochastic GPE~\cite{Blakie2008, Szigeti2009}. We assume that the system dynamics evolve on a longer timescale than the duration $\delta t$ of each probe pulse.

\section{Measurement backaction on a stable domain wall} 
For $u_2 = -0.05u_0$ we initialize a single stable domain wall and compare two measurement signals: $\mathcal{M}_{z}$ as in Fig.~\ref{Fig:1}(b) and $\mathcal{M}_{\perp}$ as in Fig.~\ref{Fig:1}(c) where $\mathcal{M}_\perp =  \sqrt{\mathcal{M}_x^2 +\mathcal{M}_y^2}$. The $\mathcal{M}_\perp$ measurement is implemented in two steps, one measurement along $x$ and one along $y$, with $\varphi \rightarrow \varphi/\sqrt{2}$ to give the same overall coupling as the single $z$ measurement; each separate measurement imparts backaction onto the condensate. The signals differ greatly; $\mathcal{M}_z$ gives a large signal everywhere atoms are present \emph{except} at the domain wall, while $\mathcal{M}_\perp$ is non-zero only within the domain wall. The domain wall width is approximately the spin healing length $\xi_s = \hbar/\sqrt{2m_anu_2}$. By fitting the $\mathcal{M}_{z}$, $\mathcal{M}_{\perp }$ to a tanh and cosh function respectively, we extract the domain wall width $\xi_{\rm w}$ and position $x_{\rm w}$ over time from the measurement signal.

\begin{figure}[t!]
\includegraphics{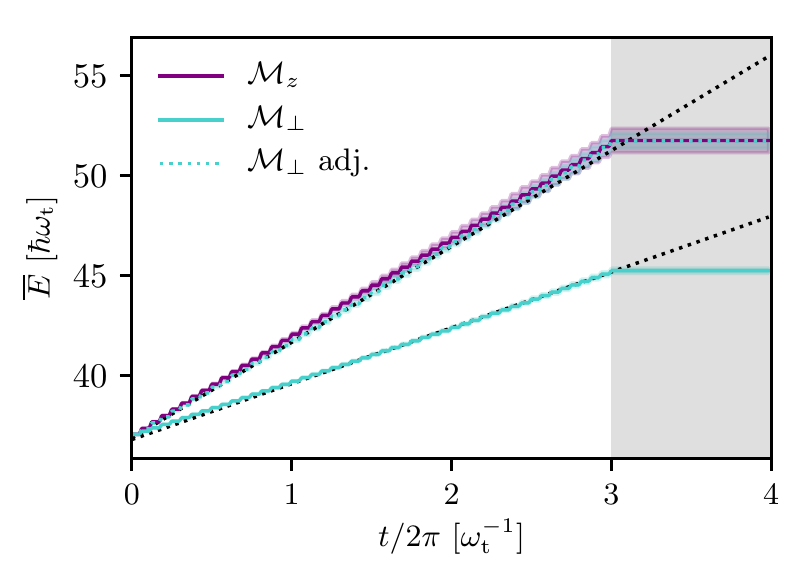}
\caption{System heating for 48 total measurements as shown by increasing energy. $\overline{E}$ for 128 trajectories is plotted and the shaded area denotes the variance. Each measurement adds energy to the system, thereby heating the system. The solid lines indicate $\varphi = 0.1$ while the dotted line indicates $\mathcal{M}_\perp$ with adjusted coupling $\varphi \approx 0.13$ which adds the same amount of energy per measurement as $\mathcal{M}_z$. The shaded area indicates time in which no measurements are taken, and energy is conserved. The dotted black lines show the analytical prediction for $\overline{E}$ from Eqn.~\eqref{Eqn:AvgE}.\label{Fig:2}}
\end{figure}

The two main effects of measurement backaction are overall system heating and domain wall diffusion. Fig.~\ref{Fig:2} summarizes heating, which we quantify in terms of the energy change per measurement $\delta E = E\left[\Psi_{|\vec{m}}\right] - E\left[\Psi\right]$, where $E$ is the GPE energy functional. From the updated amplitude in Eqn.~\eqref{Eqn:update}, we calculate
\begin{equation}
\overline{\delta E_z} \approx \frac{\varphi^2k_c}{k_M} \sum_j \left(\frac{k_c^2}{12}n_j + u_0S_{zj}^2 + u_2n_j^2\right)
\label{Eqn:AvgE}
\end{equation}
for a single measurement of $\hat{S}_z$, where $n_j$ and $S_{zj}$ denote the atom number and magnetization of the system before measurement. The first term is from the increase in kinetic energy due to measurement backaction, while the other two terms describe the change in interaction energy. For a measurement of $\hat{S}_\perp$, $\overline{\delta E_\perp}\propto \varphi^2\sum_j~u_0S_{\perp j}^2 - u_2S_{z j }^2$, which has a smaller contribution to the overall energy at equal $\varphi$ (for the domain wall) as verified numerically in Fig.~\ref{Fig:2}. Fig.~\ref{Fig:2} also shows the predicted energy increase from $\overline{\delta E_{z,\perp}}$ (dotted black lines) which agrees well with the numerical result. Adjusting the coupling for the $\mathcal{M}_\perp$ measurement to $\varphi \approx 0.13$ leads to the same energy added per measurement as for $\mathcal{M}_z$ with $\varphi = 0.1$. Thus, the choice of measurement observable affects overall system heating. 

\begin{figure}[t!]
\includegraphics{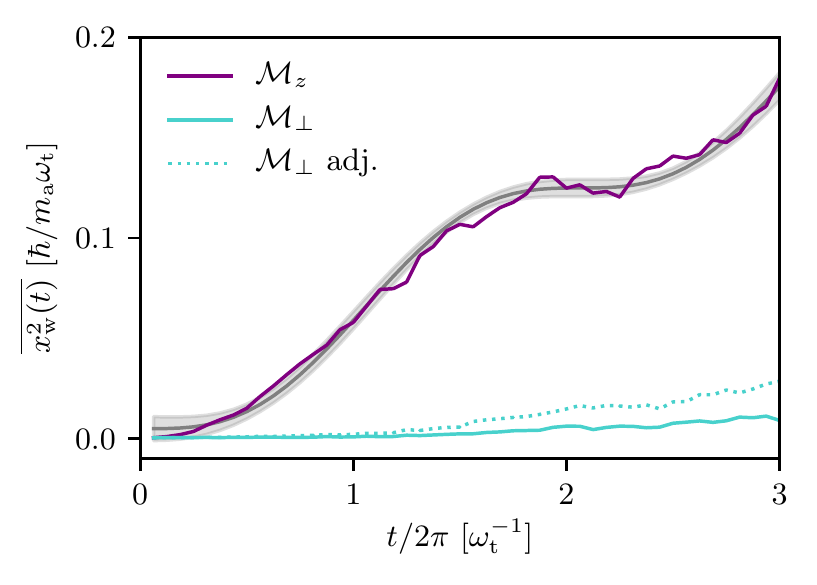}
\caption{Variance $\overline{x^2_{\rm w}(t)}$ for 128 trajectories. $\mathcal{M}_z$ shows clear diffusive behavior while $\mathcal{M}_\perp$ does not. As in Fig.~\ref{Fig:2}, the solid lines indicate $\varphi = 0.1$ while the dotted line indicates $\mathcal{M}_\perp$ with adjusted coupling $\varphi \approx 0.13$ that gives the same heating rate. The gray area shows the best-fit and uncertainty for the diffusion model in Eq.~\eqref{Eqn:diffmodel}.\label{Fig:3}}
\end{figure}

Measurement backaction also leads to diffusion effects, similar to the case of a particle coupled to a fluctuating reservoir. The domain wall is a localized, heavy object whose motion can be described by a classical Langevin theory~\cite{Risken1996, Hurst2017}. In this case the `reservoir' is the stochastic measurement backaction, which adds energy to the system after each measurement without a mechanism for dissipation.

Measurement backaction can impart noise on both the momentum $(p)$ and position $(x)$ of the domain wall. Fluctuations in $x$ correspond to measurement backaction directly changing the local spin via the Kraus operator, while fluctuations in $p$ correspond to changes in the superfluid velocity caused by density fluctuations, which create a gradient in the overall condensate phase as the system evolves in time after measurement. We account for both effects by considering a two-noise model with strengths $f_x$, $f_p$ respectively, which we assume to be anti-correlated such that $\langle f_p(t)f_x(t')\rangle = -f_xf_p\delta(t-t')$ and $\langle f_{p,x}(t)f_{p,x}(t')\rangle = f^2_{p,x}\delta(t-t')$. We quantify measurement-induced diffusion by tracking the variance, 
\begin{equation}
\overline{x^2_{\rm w}}=\frac{f_q^2+f_p^2}{2}t+\frac{f_q^2-f_p^2}{4\omega}\sin2\omega t + \frac{f_pf_q}{\omega}\sin^2\omega t + D^2_{\rm m}
\label{Eqn:diffmodel}
\end{equation}
where $\omega$ is the domain wall's oscillation frequency. The constant $D^2_{\rm m}$ accounts for initial measurement uncertainty.

Fig.~\ref{Fig:3} shows $\overline{x^2_{\rm w}(t)}$ extracted from $\mathcal{M}_z$ and $\mathcal{M}_\perp$. For $\mathcal{M}_z$ the domain wall undergoes diffusion with $\omega \approx  1.5\omega_{\rm t}$ and the noise strengths $f_{x,p} $ scale linearly with $\varphi$. In the case of $\mathcal{M}_\perp$ the measurement result stays relatively flat until $t  \approx 4\pi$, indicating that backaction due to the $\mathcal{M}_\perp$ measurement does not cause diffusion of the domain wall. At longer times, $\overline{x^2_{\rm w}(t)}$ does begin to increase, which we attribute to overall heating rather than measurement backaction. This shows that measurement backaction due to $\mathcal{M}_z$ is more disruptive to the domain wall because each measurement imparts backaction noise across the whole atom cloud, whereas the backaction for the $\mathcal{M}_\perp$ measurement occurs only at the domain wall center and does not affect the density away from the domain wall.

\section{Feedback-stabilized domain wall} 
We now turn to creating and stabilizing a domain wall using a measurement of $\hat{S}_z$ followed by classical feedback. We start with a condensate with $u_2 = 0.05u_0$ which forms a uniform condensate polarized in $xy$ (easy-plane) with $\langle \hat{S}_{z} \rangle = 0$, where in the closed system a domain wall is not energetically favorable~\cite{Note1}. We derive a feedback signal
\begin{equation}
w = \frac{1}{N}\sum_j \mathrm{sgn}(x_j) \mathcal{M}_{z}(x_j)
\label{Eqn:Feedback}
\end{equation}
from each measurement $\mathcal{M}_z$, where on average $\bar{w}=0$ for a uniformly easy-axis or easy-plane polarized phase and approaches $\bar{w} = \pm 1$ for a domain wall centered at $x=0$; the sign identifies the orientation. For example, the domain wall signal in Fig.~\ref{Fig:1}(b) has $w = -0.99$. We then apply a magnetic field gradient $V_{z}(x_j) = gw\tau^z x_j$ with strength proportional to $w$ and gain $g$.

\begin{figure}[t!]
\includegraphics{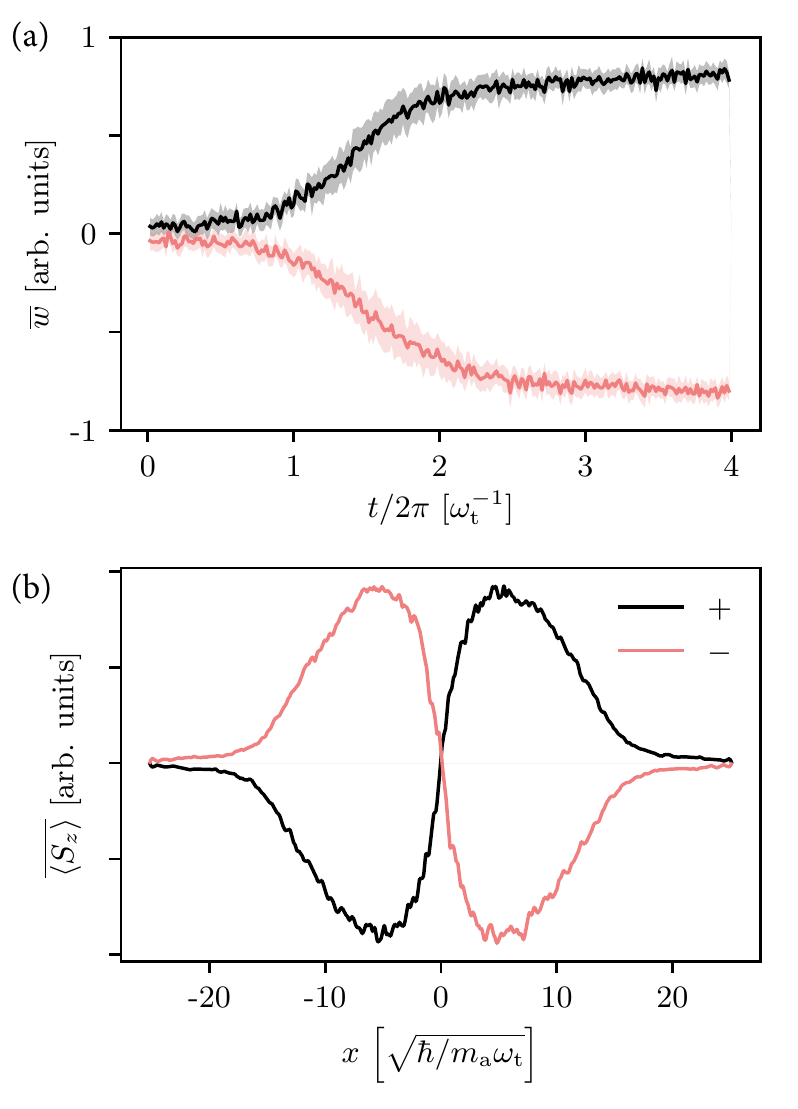}
\caption{(a)~Domain wall signal $w$ for $\varphi = 0.01$ and $g = -5$. The solid lines are the average over trajectories in the $+$ or $-$ branch and the semi-transparent area indicates the variance. A domain wall is formed within 3 trap periods with the orientation spontaneously determined based on the first few measurements of the system.~(b)~Final value of $\langle S_z \rangle$ averaged over all trajectories for the $+$ (black) and $-$ (pink) domain wall orientations; the variance is same as the linewidth. The internal spin-dependent interaction parameter is $u_2 =  0.05u_0$.
\label{Fig:4}}
\end{figure}

Figure~\ref{Fig:4} summarizes the results of feedback. Initially the condensate is spin-unpolarized and $w$ randomly fluctuates about zero. After a few measurements the sign stabilizes and $|w|$ increases, signifying domain wall formation with a stable orientation as shown by the two branches of $\overline{w}$ in Fig.~\ref{Fig:4}(a). The average $\overline{w}$ for $\pm$ orientations is calculated by binning the trajectories by the sign of $w$ at the final timestep. Here, the band indicates the variance of all trajectories on each branch. The process is nearly symmetric; out of 256 total trajectories, 122 evolved to  the `$+$' orientation with $\overline{w} = 0.78$ and 134 to the `$-$' orientation with $\overline{w} = -0.8$. This bistability is reminiscent of spontaneous symmetry breaking in ferromagnets, but here quantum measurement and feedback ``spontaneously" broke the initial symmetry~\cite{Note1}.

In Fig.~\ref{Fig:4}(b) we show $\overline{\langle S_{z}\rangle}$ for each final orientation, which clearly shows the presence of a domain wall. This is reminiscent of the ground state of a two-component BEC in the immiscible regime with $u_2 < 0$,  even though the internal interaction parameter is $u_2 = 0.05u_0$. This shows that measurement and feedback can be used to stabilize phases that would not be stable in equilibrium. However, our demonstration protocol is not quite the same as tuning interactions locally because $w$ in Eq.~\eqref{Eqn:Feedback} is not spatially dependent. This type of feedback could not lead to the formation of multiple domains, which happens when $u_2$ is rapidly quenched~\cite{De2014, Hofmann2014}.

\section{Outlook} 
We outlined a new way to dynamically create stable spin textures in cold gases that is directly applicable to other systems such as Fermi gases or atoms in optical lattices. Repeated weak measurements eventually heat the system, which can be mitigated in experiment by evaporation, or even by suitable local feedback~\cite{Hush2013, Wade2016}. This work poses new questions such as: Can spatially dependent feedback lead to an effective description with changed interaction parameters? How can feedback maximally control heating? Future research could address these questions using other types of feedback or different measurement observables. Finally, additional sources of noise in measurements could make feedback less efficient. Expanding the theory to include detector inefficiencies and technical noise is an important step toward implementing our proposal, and will be addressed in future work.

\begin{acknowledgments}
This work was partially supported by the Air Force Office of Scientific Research's Quantum Matter MURI, NIST, and NSF (through the Physics Frontier Center at the JQI). HMH acknowledges the support of the NIST/NRC postdoctoral program.
\end{acknowledgments}
\bibliography{main}
\begin{widetext}
\section*{Supplementary Material: Measurement induced dynamics and stabilization of spinor condensate domain walls}

\section*{A. Measurement Model Details}

Just prior to measurement, the system and light pulse at $x_j$ each evolve together under $\hat{U}_{\vec{r}j} = \exp\left[-i\delta t \hat{H}_{\vec{r}j}^{\rm SR}/\hbar\right]$ where 
\begin{equation}
\hat{H}^{\rm SR}_{\vec{r}j} =  \frac{\hbar\gamma}{c\delta t}\hat{S}_{\textbf{r} j} \otimes \hat{n}_j.
\label{Eqn:HSR2}
\end{equation}
We take the probe field amplitude to be strong enough that the light is still nearly a coherent state after interacting with the atoms such that $\hat{a}_j \approx \langle \hat{a}_j\rangle +\delta\hat{a}_j$. To first order in $\delta\hat{a}_j$, we then have $\hat{n}_j \approx \sqrt{2}\mathrm{Re}\alpha\hat{X}_j + \sqrt{2}\mathrm{Im}\alpha\hat{Q}_j - |\alpha|^2$ where $\hat{X}_j = (\hat{a}_j +\hat{a}_j^\dagger)/\sqrt{2}$ and $\hat{Q}_j =  (\hat{a}_j-\hat{a}_j^\dagger)/\sqrt{2}i$ are quadrature variables with $\left[\hat{X}_j,\hat{Q}_{j'}\right] = i\delta_{jj'}$. Thus, up to a global phase the evolution operator is $\hat{U}_{\vec{r}} = \exp\left[-i\sum_j\hat{S}_{\vec{r}j}\otimes\left(\varphi^x\hat{X}_j + \varphi^q\hat{Q}_j\right)\right]$ with couplings $\varphi^x= \sqrt{2}\gamma|\alpha|\cos\phi/c$, $\varphi^q =\sqrt{2}\gamma|\alpha|\sin\phi/c$. We then set $\phi = \pi/2$ which gives $\varphi^x = 0$ and $\varphi^q \rightarrow \varphi = \sqrt{2}\gamma|\alpha|/c$. The beam is homodyne detected on an array of detectors; during homodyne detection the reservoir state is strongly measured in the eigenbasis of the $\hat{X}_j$ operators with eigenvalues $\hat{X}_j|m_j\rangle = m_j|m_j\rangle$. The reservoir state $|\alpha\rangle$ is assumed to be Gaussian over the $|m_j\rangle$ states (suitable for a coherent state of light), leading to Gaussian-distributed measurement outcomes $m_j$. Thus, the measurement outcome for the full detector array is a vector $\vec{m}(x_j) = (m_1, m_2, \ldots, m_j)$. When coupled to the quantum system, $\hat{U}_{\vec{r}}$ locally shifts the $|m_j\rangle$ states by $\varphi \langle \hat{S}_{\vec{r} j}\rangle$. The system wavefunction after measurement is $|\Psi_{|\vec{m}}\rangle = \hat{\mathcal{K}}_{\vec{r}|\vec{m}}|\Psi\rangle$ where $\hat{\mathcal{K}}_{\vec{r}|\vec{m}} = \langle \vec{m}|\hat{U}_{\vec{r}}(\delta t)|\alpha\rangle$ is a Kraus operator corresponding to a specific measurement outcome $\vec{m}(x_j)$ and $|\Psi\rangle$ is the system state before measurement. We present the functional form of $\hat{\mathcal{K}}_{\vec{r}|\vec{m}}$ in the main text by expanding the formal expression to $\mathcal{O}(\varphi^2)$. 

\section*{B. Simulation Parameters}
For each simulation the internal dynamics of the system (Eq. (4) in the main text) were modeled via a Gross-Pitaevskii equation (GPE) using the split-step integration method in Ref.~\cite{Symes2016}. First, we found the ground state of the GPE via  imaginary time $t\rightarrow -i\tau$, using the strong convergence critertion in Ref.~\cite{Antoine2014} to test for convergence. Then, we studied the effect of measurement by running the GPE in real time to account for internal dynamics, and applying the Kraus operator (Eq. (5) in the main text) each time we `measured' the system. We studied the effect of measurement backaction in the regime where domain walls are stable ($u_2/u_0 < 0$) and  we studied measurement and feedback in the regime where they are unstable ($u_2/u_0 > 0$). The number of particles was fixed to $N = 10^4$, the time increment was $dt = 3.8\times 10^{-4}\omega_{\rm t}^{-1}$, and the spatial increment was $\Delta x \approx 0.02$. 

In the \emph{measurement backaction} section of the main manuscript, we study the measurement backaction on the BEC in the regime where domain walls are stable. These simulations (Figs. 1. 2. and 3 in the main text) were run with $u_2/u_0 = -0.05$, $u_0 = 0.1\Delta x$ and the initial condition is given in Fig.~\ref{Fig:S2}
\begin{figure}[H]
\centering
\includegraphics{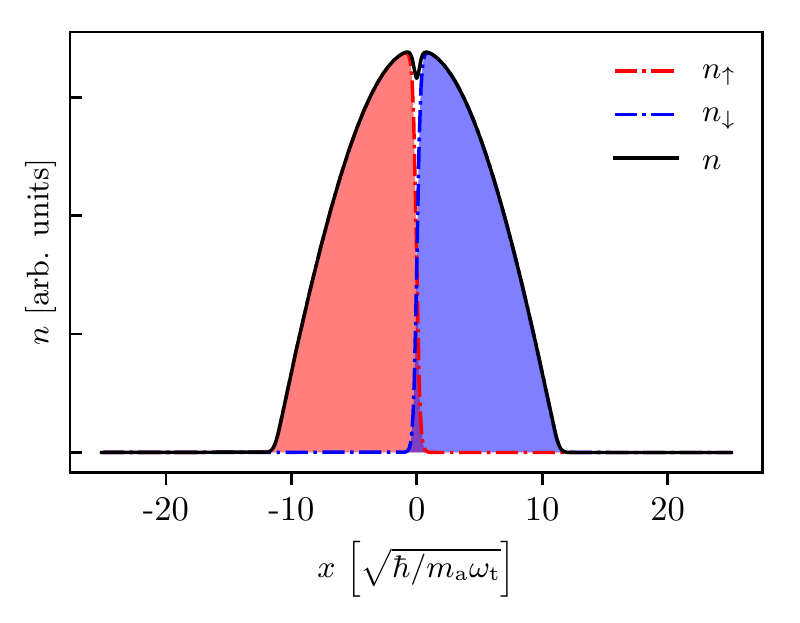}
\caption{Computed ground state system in the immiscible regime with a single domain wall. Initital condition for the `measurement backaction' section.\label{Fig:S2}}
\end{figure}
In the \emph{feedback-stabilized domain wall} section of the paper we started the measurement and feedback from the ground state of a spin-unpolarized system. These simulations (Fig. 4 in the main text) were run with $u_2/u_0 = 0.05$, $u_0 = 0.1\Delta x$ and the initial condition is given in Fig.~\ref{Fig:S3}.
\begin{figure}[H]
\centering
\includegraphics{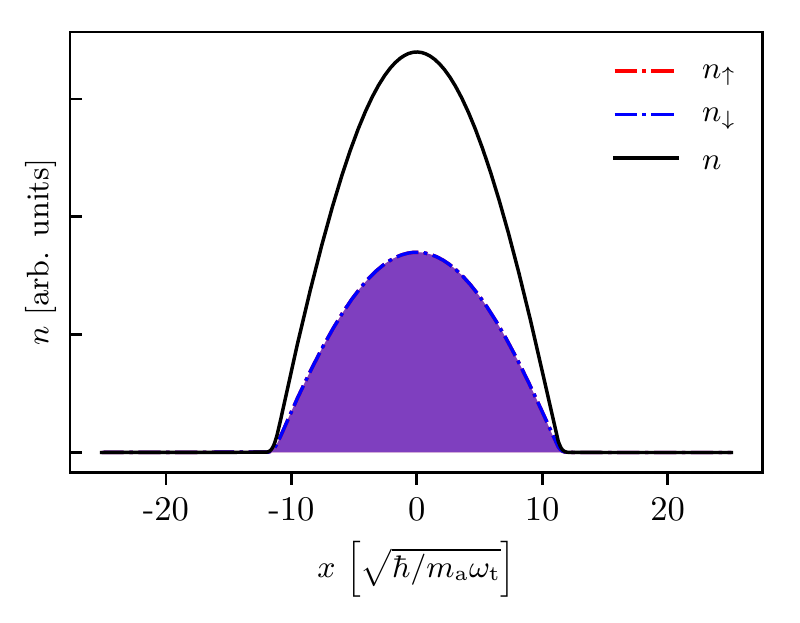}
\caption{Computed ground state system in the miscible regime with equal, evenly distributed spin population. Initial condition for the `feedback' section.\label{Fig:S3}}
\end{figure}

\section*{C. Behavior of Individual Trajectories Under Feedback}
\begin{figure}[H]
\centering
\includegraphics{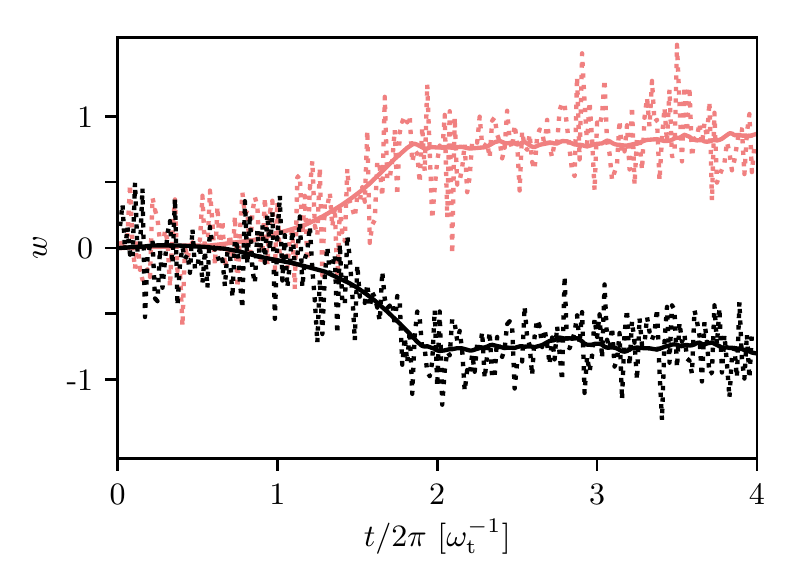}
\caption{Example of two individual system trajectories under measurement and feedback for $\varphi = 0.01$ and $g = -5$. The dotted lines indicate the measurement result (with measurement noise) and the solid lines are calculated using the wavefunction only. Notice at short times ($t/2\pi < 1$) the measurement trajectories oscillate around zero and the solid lines change sign before stabilizing around $w \approx \pm 1$. \label{Fig:S1}}
\end{figure}

Under measurement and feedback, individual system trajectories show signatures of spontaneous symmetry breaking. The sign of the feedback signal $w$ (defined in the main text) determines the orientation of the domain wall. Fig.~\ref{Fig:S1} shows the evolution of $w$ for two system trajectories under measurement and feedback, showing that the sign of $w$ does not stabilize for $t/2\pi < 1$. The average $\overline{w}$ for $\pm$ orientations is calculated by binning the trajectories based on the sign of $w$ at the final timestep. 
\end{widetext}
\end{document}